\documentclass[a4paper,twoside]{article}

\usepackage{epsfig}
\usepackage{hyperref}
\usepackage{subfigure}
\usepackage{calc}
\usepackage{amssymb}
\usepackage{amstext}
\usepackage{amsmath}
\usepackage{amsthm}
\usepackage{multicol}
\usepackage{pslatex}
\usepackage{apalike}
\usepackage{SCITEPRESS}     
\usepackage{amsmath}
\usepackage{amsfonts}
\usepackage{amssymb}
\usepackage{colortbl} 

\begin{document}

\title{A Simple and Correct Even-Odd Algorithm for the Point-in-Polygon Problem for Complex Polygons}

\author{\authorname{Michael Galetzka\sup{1} and Patrick Glauner\sup{2}}
\affiliation{\sup{1}Disy Informationssysteme GmbH, Ludwig-Erhard-Allee 6, 76131 Karlsruhe, Germany}
\affiliation{\sup{2}Interdisciplinary Centre for Security, Reliability and Trust, University of Luxembourg, 4 rue Alphonse Weicker,\\ 2721 Luxembourg, Luxembourg}
\email{galetzka.michael@gmail.com, patrick.glauner@uni.lu}
}


\abstract{Determining if a point is in a polygon or not is used by a lot of applications in computer graphics, computer games and geoinformatics. Implementing this check is error-prone since there are many special cases to be considered. This holds true in particular for complex polygons whose edges intersect each other creating holes. In this paper we present a simple even-odd algorithm to solve this problem for complex polygons in linear time and prove its correctness for all possible points and polygons. We furthermore provide examples and implementation notes for this algorithm.}

\onecolumn \maketitle \normalsize \vfill

\section{\uppercase{Introduction}}
\label{sec:introduction}

\begin{figure}[h]
	\begin{center}
  		\includegraphics[scale=0.3]{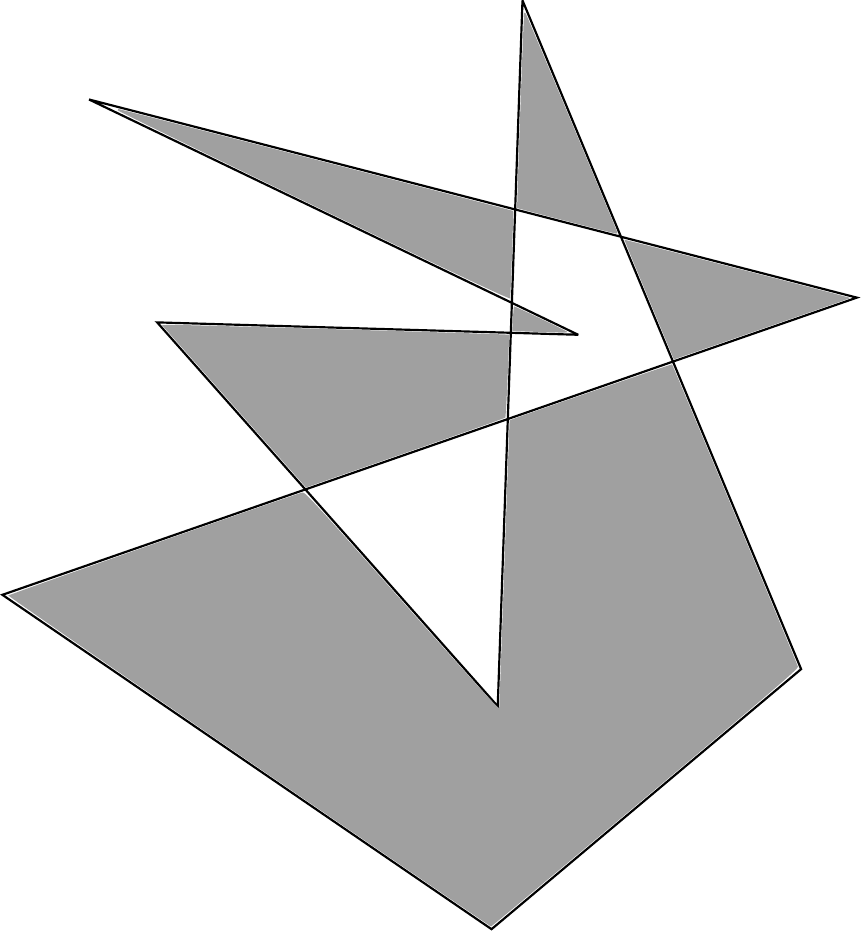}
	\end{center}
	\caption{A self-intersecting polygon.}
	\label{IntersectingPolygon}
\end{figure}

At a first glance, the point-in-polygon problem seems to be a rather simple problem of geometry: given an arbitrary point $Q$ and a closed polygon $P$, the question is whether the point lies inside or outside the polygon.
There exist different algorithms to solve this problem, such as a cell-based algorithm \cite{Zalik:20011135}, the winding number algorithm \cite{Hormann:2001131} or the even-odd algorithm \cite{Foley:1990} that is used in this paper. The problem is not as trivial as it seems to be if the edges of the polygon can intersect other edges as seen in Figure~\ref{IntersectingPolygon}. This kind of polygon is also often called a complex polygon since it can contain "holes".

A lot of special cases have to be considered (e.g. if the point lies on an edge) and most of the existing even-odd algorithms either fail one or more of these special cases or have to implement some kind of workaround \cite{Schirra:2008}. The rest of this paper is organized as follows. In Section~\ref{sec:alg}, we present an even-odd algorithm and prove its correctness for all possible points and polygons with no special cases to be considered. In Section~\ref{sec:example}, we apply this algorithm to an example complex polygon. We then provide implementation notes in Section~\ref{sec:im}. Section~\ref{sec:con} summarizes this work.

\section{\uppercase{The even-odd algorithm}}
\label{sec:alg}
The basic even-odd algorithm itself is fairly simple: cast an infinite ray from the point in question and count how many edges of the polygon the ray intersects \cite{Foley:1990}.

\paragraph{Definition}
Let $P$ be a polygon with $n$ vertices $P_{1}, ..., P_{n-1}, P_{n}$ where the vertices are sorted in such a way that there exists an edge between $P_{i}$ and $P_{i+1}$ for every index $1\le i<n$ and between $P_{1}$ and $P_{n}$.
The algorithm computes if the arbitrary point $Q$ lies either $inside$ or $outside$ of $P$.

\paragraph{The steps of the algorithm}

To ease the presentation of the steps of the algorithm, it defines its own coordinate system with $(0|0)=Q$ by moving the polygon. This translation is only for illustration purposes. The algorithm then uses the positive x-axis as ray to calculate intersections with the edges of $P$. So the start vertex of the ray is $(0|0)$ and the end vertex is $(x_{max}|0)$ where $x_{max}$ is an x value greater than that of any of the vertices of $P$.
\begin{enumerate}
\item First it is determined if $Q$ is equal to any of the vertices of $P$ or lies on any of the edges connecting the vertices. If so the result is $inside$.
\item A vertex $P_{s}$ that does not lie on the x-axis is searched in the set of vertices of $P$. If no such vertex can be found the result is $outside$.
\item Set $i$ to 1. Beginning from that vertex $P_{s}$ the following steps are repeated until all vertices of $P$ have been visited:
	\begin{enumerate}
	\item The index $s$ is increased to $s+i$ until the next vertex $P_{s+i}$ not lying on the x-axis is found. If the index $s+i>n$ then $i$ is set to $-s$ and the search is continued.
	\item Depending on the course of step (a) one of the following steps is taken:
		\begin{enumerate}
		\item \textit{No vertex has been skipped}: the line segment from $P_{s}$ to $P_{s+i}$ is intersected with the positive x-axis.
		\item \textit{At least one vertex with a positive x-value has been skipped}: the line segment from $P_{s}$ to $P_{s+i}$ is intersected with the complete x-axis.
		\item \textit{At least one vertex with a negative x-value has been skipped}: nothing is done.
		\end{enumerate}
	\item $P_{s+i}$ is the starting vertex for the next iteration.
	\end{enumerate}
\item If the count of intersections with the x-axis is even then the result is $outside$, if it is odd the result is $inside$.
\end{enumerate}

\paragraph{Proof of correctness} To prove that the oven-odd algorithm in general is correct one can generalize it to the Jordan Curve Theorem and prove its correctness \cite{Tverberg:01011980}. What has to be proven is that the given algorithm is in fact a correct even-odd algorithm, meaning that the count of edges is correct under all circumstances.

A lot of challenges emerge when trying to intersect two line segments and one of the segments has one or two of its vertices lying on the other segment. If the count of intersections with the x-axis are to be counted correctly, then each edge of $P$ must either clearly intersect the x-axis or not intersect it at all. There must be no case where the starting or end vertex of a line segment lies on the line segment it should be intersected with.

The x-axis is the first line segment to look at, since it is part of all the intersections. The start vertex of the x-axis, namely $Q$, is guaranteed not to lie on any edge or to be equal to any vertex of $P$. If this was the case, then the first step of the algorithm would have already returned the correct result. The end vertex is guaranteed to have an x value greater than any of the vertices of $P$, so no vertex of $P$ can be equal to it and no edge of $P$ can contain it.

\begin{figure}[h]
	\begin{center}
  		\includegraphics[scale=0.6]{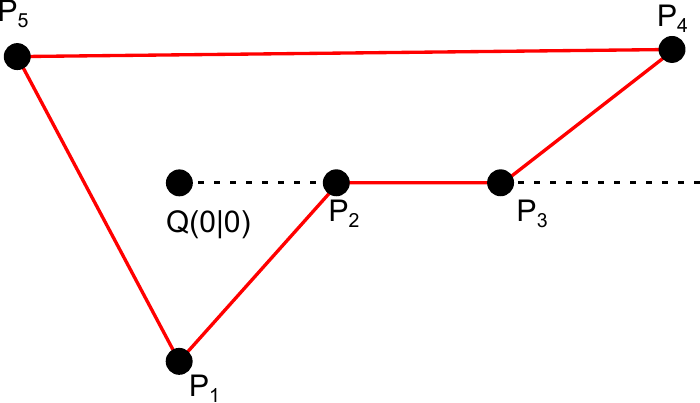}
	\end{center}
	\caption{One of the edges of $P$ lies on the x-axis.}
	\label{edgeOnAxisProblem}
\end{figure}

Of course there still exists the challenge that one or more vertices of $P$ lie on the x-axis as seen in Figure~\ref{edgeOnAxisProblem}.

\begin{figure}[h]
	\begin{center}
  		\includegraphics[scale=0.6]{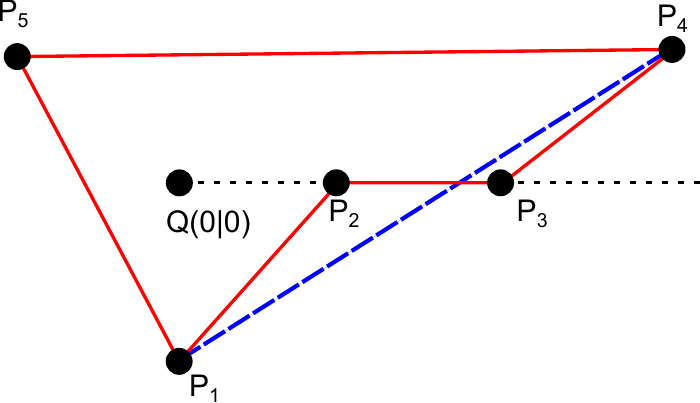}
	\end{center}
	\caption{A new auxiliary edge has been created.}
	\label{edgeOnAxisSolution}
\end{figure}

The algorithm deals with this kind of challenge by ignoring the vertices lying on the x-axis and stepping over them when trying to find an edge to intersect. It then creates a new auxiliary edge that it can intersect safely. So starting for example at vertex $P_{1}$ it would ignore $P_{2}$ and $P_{3}$ and then create a new edge between $P_{1}$ and $P_{4}$ as shown in Figure~\ref{edgeOnAxisSolution}.

At this point it is clear that none of the edges used to calculate the count of intersections with the x-axis has a vertex on the x-axis and is either clearly intersecting it or not. What has to be shown is that all created auxiliary edges are correct substitutes to calculate the intersections with the x-axis.

\begin{figure}[h]
	\begin{center}
  		\includegraphics[scale=0.6]{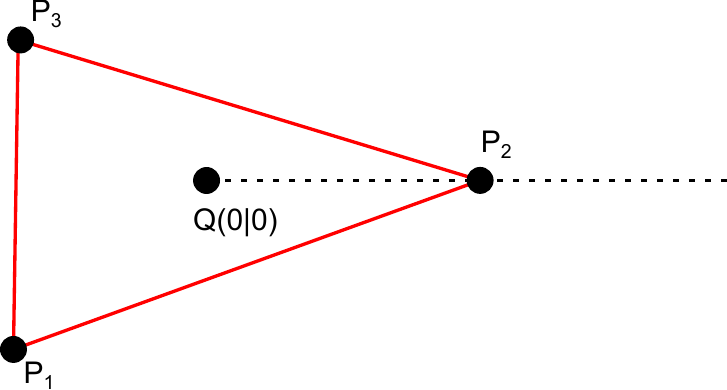}
	\end{center}
	\caption{The auxiliary edge between $P_{1}$ and $P_{3}$ does not intersect the positive x-axis.}
	\label{auxiliaryEdgeProblem}
\end{figure}

Indeed they are not a correct substitute to intersect the positive x-axis, as can be seen in Figure~\ref{auxiliaryEdgeProblem}. Here the auxiliary edge would be the same as the edge between $P_{1}$ and $P_{3}$ and this edge does clearly not intersect the positive x-axis. So the total count of intersected edges of the polygon shown in Figure~\ref{auxiliaryEdgeProblem} would be zero - which is obviously wrong.

The algorithm actually deals with this challenge in step $3.b$, by extending the ray in that special case where a vertex lying on the positive x-axis has been skipped. The new ray is then the complete x-axis and not only the positive part of it. It can be seen that this would create the desired result in the example of Figure~\ref{auxiliaryEdgeProblem}, because the total count of intersected edges would then be one.

The question remains if this method always yields the correct result under all possible circumstances. If the skipped vertex lies on the negative x-axis then the auxiliary line will not be considered for intersection, since none of the original edges could have intersected the positive x-axis. So all cases that have to be looked at involve one or more vertices on the positive x-axis. The skipped vertices can never change from the positive to the negative x-axis since this would have been caught in the first step of the algorithm.

So, all auxiliary edges that intersect the x-axis are created from the following order of vertices: $P_{u}$ which does not lie on the x-axis, $P_{v}$\footnote{Of course there could be more than just one vertex on the x-axis but they are all skipped and the same auxiliary edge is created no matter how many additional vertices are on the x-axis.} which does lie on the positive x-axis and $P_{w}$ which does not lie on the x-axis. Each of the vertices $P_{u}$ and $P_{w}$ can lie in one of the four quadrants around the point $Q$ and therefore there exist $4\times 4=16$ different versions of the auxiliary edge that have to be considered. This number can be further reduced to $10$ because six of these versions are created by switching start and end vertex and do not have to be considered as a separate case.

All possibilities of the locations of $P_{u}$ and $P_{w}$ are shown in Table~\ref{auxiliaryTable} as well as the desired intersection count and the actual intersection count of the algorithm. It is clear by looking at the table that the algorithm satisfies the desired result for each possible scenario. Therefore the auxiliary line is indeed a correct substitute for the original edges.

This leads to the conclusion that the algorithm provided can correctly determine if there is an even or an odd number of intersections with the polygon.

\hfill $\square$

\paragraph{Time complexity} All $n$ vertices of the polygon are visited once during the translation and the first three steps of the algorithm. All other computations like step four can be completed within constant time. 
Therefore the time complexity for this algorithm is $\mathcal O(n)$.

\hfill $\square$ 

\section{\uppercase{Example}}
\label{sec:example}
\begin{figure}[h]
	\begin{center}
  		\includegraphics[scale=0.6]{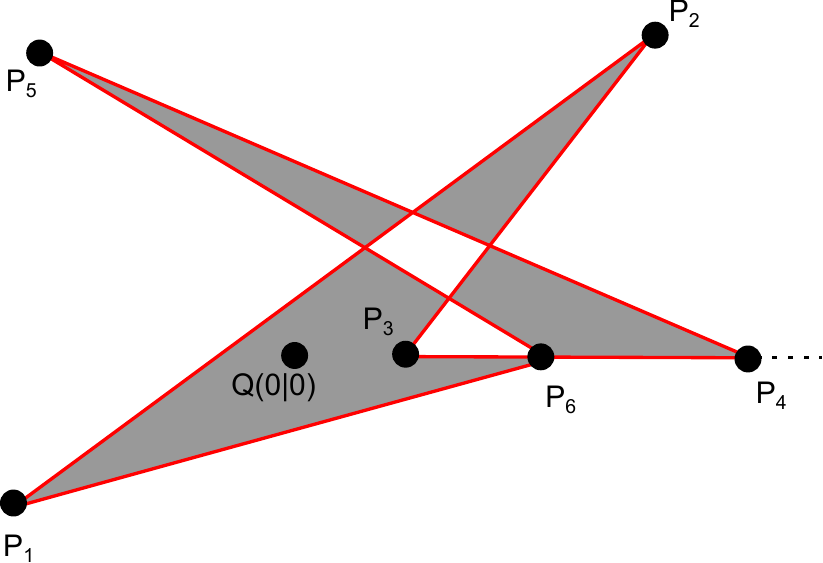}
	\end{center}
	\caption{Sample complex polygon $P$.}
	\label{ExamplePolygon}
\end{figure}

The algorithm is applied to the sample complex polygon $P$ in Figure~\ref{ExamplePolygon}.

\begin{enumerate}
\item $Q$ lies neither on any vertex nor edge.
\item The start vertex $P_{s}$ is  $P_{1}$ in this example.
\item All intersection and substitution steps can be found in Table~\ref{exampleTable}.
\item Since the count of intersections with the x-axis is odd the result is $inside$.
\setlength{\tabcolsep}{1pt}
\begin{table}[h]
\caption{Intersection and substitution steps.}
\begin{tabular}{|c|c|c|c|c|}
\hline 
$P_{u}$ & $P_{v}$ & $P_{w}$ & x-axis for intersection operation & Inters. \\ 
\hline 
$P_{1}$ & $-$ & $P_{2}$ & positive & no \\ 
\hline
$P_{2}$ & $(P_{3}, P_{4})$ & $P_{5}$ & complete & no \\ 
\hline 
$P_{5}$ & $P_{6}$ & $P_{1}$ & complete & yes \\ 
\hline 
\end{tabular} 
  \label{exampleTable}
\end{table}
\end{enumerate}

\section{\uppercase{Implementation}}
\label{sec:im}
The translation of $P$ can be done together with the first step of the algorithm by moving the vertices by the negative x- and y-values of $Q$. The geometric helper functions to intersect line segments can be found in \cite{Stueker:1999} which are an improved version of \cite{Sedgewick:1992}. Our reference implementation of the even-odd algorithm for complex polygons presented in this paper is available as open source: \url{http://github.com/pglauner/point_in_polygon}.

\section{\uppercase{Conclusions}}
\label{sec:con}

The implementation of point-in-polygon algorithms for complex polygons is error-prone since there are many special cases to be considered. We presented a correct even-odd algorithm that prevents special cases by substituting a sequence of edges with an auxiliary edge.
The correctness of the algorithm has been proven for all possible polygons, including complex polygons.
The algorithm is also time-efficient since it is $\mathcal O(n)$ for polygons with $n$ vertices.

\bibliographystyle{apalike}

\newpage
\section*{\uppercase{Appendix}}

\setlength{\tabcolsep}{0pt}

\begin{table}[h]
\caption{Possible cases when creating an auxiliary line (blue). The values $q1$ to $q4$ correspond to the quadrants of a cartesian coordinate system ($q1: x>0\wedge y>0,q2: x<0\wedge y>0, ...$). Rows that are just gained by switching the vertices $P_{u}$ and $P_{w}$ are omitted due to symmetry.}
\begin{tabular}{|c|c|c|c|c|}
\hline 
$P_{u}$ & $P_{w}$ & Example & Desired count & Inters. \\ 
\hline 
$q1$ & $q1$ & \includegraphics[scale=0.295]{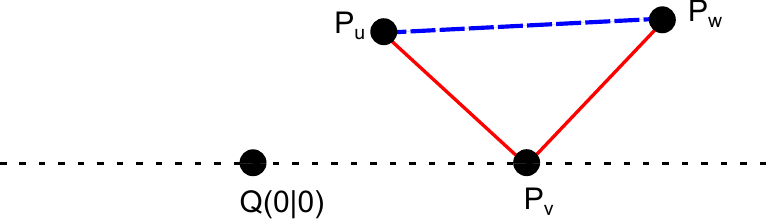} & even & no \\ 
\hline 
$q1$ & $q2$ & \includegraphics[scale=0.295]{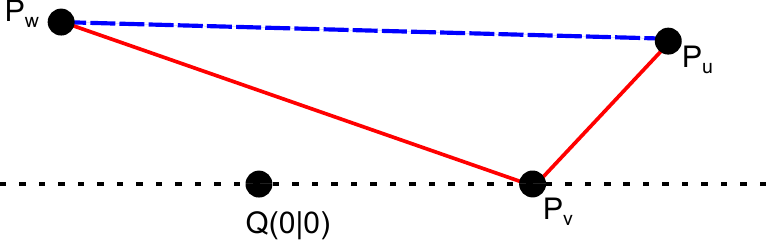} & even & no \\ 
\hline 
$q1$ & $q3$ & \includegraphics[scale=0.295]{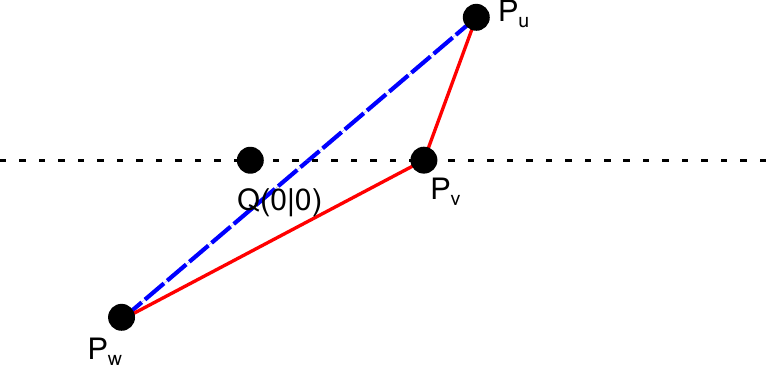} & odd & yes \\ 
\hline 
$q1$ & $q4$ & \includegraphics[scale=0.295]{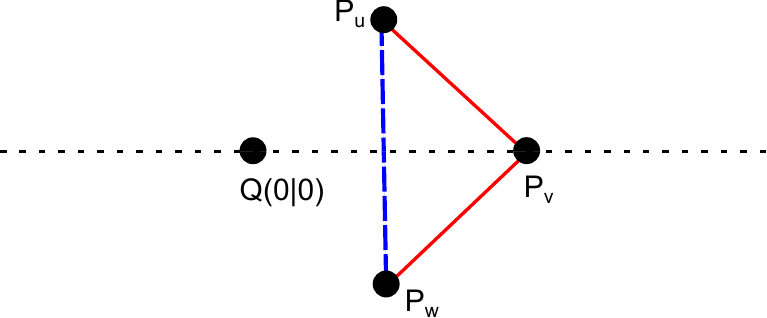} & odd & yes \\ 
\hline 
$q2$ & $q2$ & \includegraphics[scale=0.295]{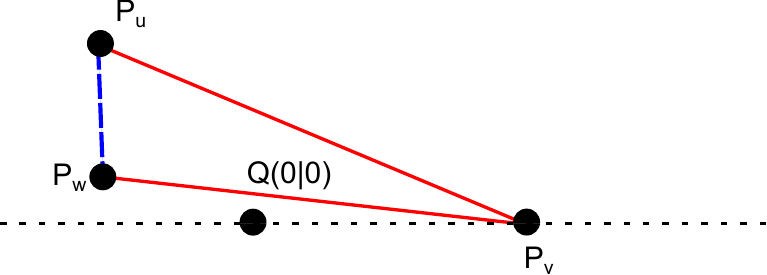} & even & no \\ 
\hline 
$q2$ & $q3$ & \includegraphics[scale=0.295]{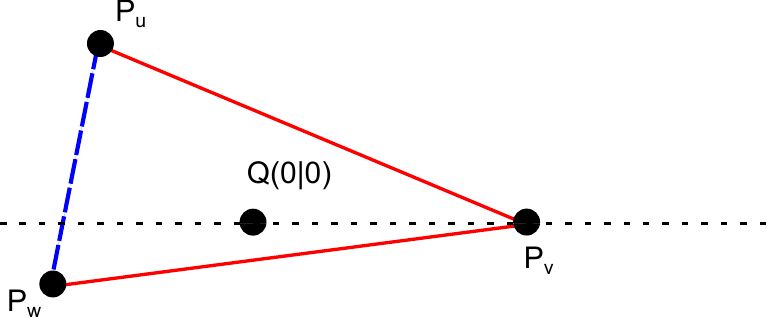} & odd & yes \\ 
\hline 
$q2$ & $q4$ & \includegraphics[scale=0.295]{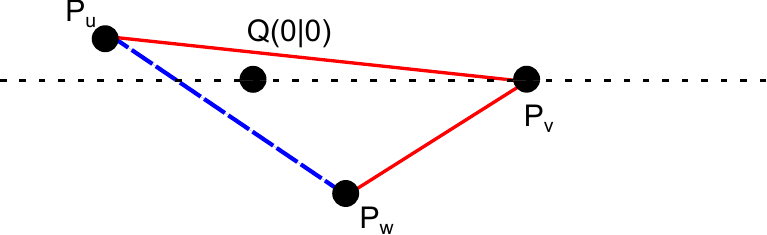} & odd & yes \\ 
\hline 
$q3$ & $q3$ & \includegraphics[scale=0.295]{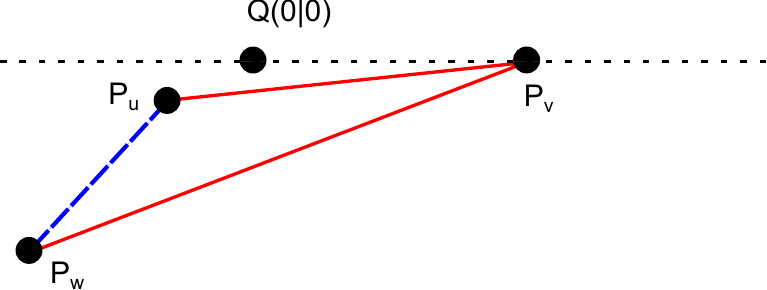} & even & no \\ 
\hline 
$q3$ & $q4$ & \includegraphics[scale=0.295]{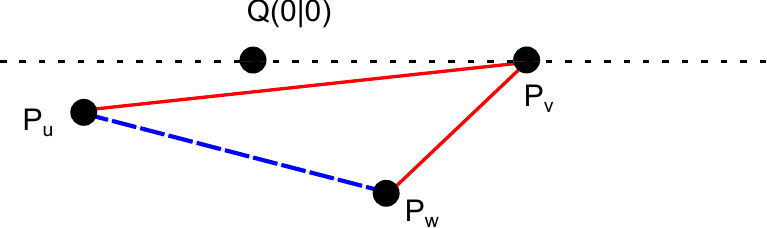} & even & no \\ 
\hline 
$q4$ & $q4$ & \includegraphics[scale=0.295]{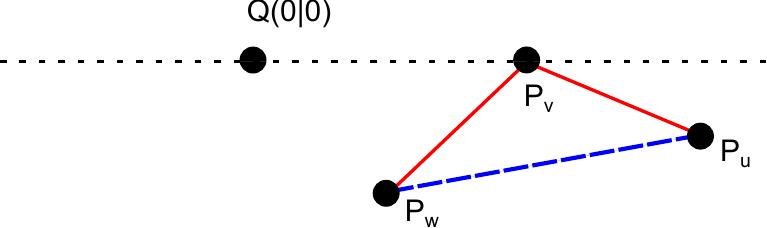} & even & no \\ 
\hline 
\end{tabular} 
  \label{auxiliaryTable}
\end{table}

\vfill
\end{document}